
\magnification 1200
\baselineskip=6truemm

\def\bbar{\overline{b}}
\def\dbar{\overline{d}}
\def\cbar{\overline{c}}
\def\ubar{\overline{u}}
\def\Kbar{\overline{K}}
\def\to{\rightarrow}

\line{\hfil TECHNION-PH-95-7}
\line{\hfil February 1995}
\null\vskip 1truecm
\centerline{\bf Large CP Asymmetries in $B^{\pm}\to\eta_c(\chi_{c0})\pi^{\pm}$
from $\eta_c(\chi_{c0})$ Width} \vskip 2truecm
\centerline{Gad Eilam, Michael Gronau and Roberto R. Mendel $^*$}
\vskip 1truecm
\centerline{Department of Physics}
\centerline{Technion -- Israel Institute of Technology}
\centerline{32000 Haifa, Israel}

\vskip 2truecm
\centerline{\it ABSTRACT}
\baselineskip=9truemm
\noindent
We study CP asymmetries in
$B^{\pm}\to h\pi^{\pm}$ decays, where the hadronic states $h=\rho\rho,
K\Kbar\pi$,  $\pi^+\pi^-K^+K^-$, etc., and $h=\pi^+\pi^-, K^+K^-,
2(\pi^+\pi^-)$, etc., are taken on the resonances $\eta_c$ and $\chi_{c0}$,
respectively.
The relatively large $\eta_c$ and $\chi_{c0}$ decay widths, of
about 10$-$15 MeV, provide the necessary absorptive phase in the interference
between the resonance (going through $b\to c\cbar d$) and the background
(through $b\to u\ubar d$) contributions to the amplitude.
Large asymmetries of order 10$\%$ or more are likely in some modes.

\vskip 4truecm
$^*$ On Sabbatical leave from the Department of Applied Mathematics, University
of Western Ontario, London, Ontario, Canada

\vfill\eject

In the Standard Model one expects CP violation to show up in $B$ decays in a
variety of ways [1]. The most promising CP asymmetry seems to occur in neutral
$B$ decays, such as $B^0\to\psi K_S$, of the
type $b\to c\cbar s$. The large time-dependent asymmetry of this processes is
given cleanly by the CKM phase $\beta=-{\rm arg}(V_{td})$ (we use the standard
convention [2]). CP asymmetries in
charged $B$ decays (which are self-tagging)
are in principle easier to measure. They are, however, harder to
calculate, since they
usually depend on hadronic matrix elements of quark operators and on unknown
final state strong interaction phases. Aside from upper limits [3],
there is no evidence for final state phases in $B$ decays and it is
often argued that these phases are small because of the high $B$ mass.
As an example, for the
Cabibbo-suppressed process $B^-\to\psi\pi^-$ [4] (going dominantly
through $b\to c\cbar d$) a model was used [5] to
describe the rescattering from intermediate
hadronic states formed by $c\cbar d\ubar$ and $u\ubar d\ubar$
to the final $\psi\pi^-$ state. A small rescattering phase was
found, leading to an asymmetry at a percent level.

In the present Letter we will show that
one way to overcome small final state interaction phases is to look for
$B$ decays that go through wide $c\cbar$ resonances, where the
resonance width provides the necessary phase.
We will study charged $B$ decay processes dominated by
$b\to c\cbar d$, in which the $c\cbar$ pair forms one of the two known wide
spin-zero states, $\eta_c$ and $\chi_{c0}$. These states may be
identified by their hadronic decay modes, e.g. $\rho\rho, K\Kbar\pi,
\pi^+\pi^-K^+K^-$ and $2(\pi^+\pi^-), \pi^+\pi^-K^+K^-$, respectively, for
which
typical branching ratios of  a few  percent have been measured, or by
$\pi^+\pi^-, K^+K^-$ into which $\chi_{c0}$ decays at a percent level [2]. The
relatively large $\eta_c$ and $\chi_{c0}$ decay widths, of about 10$-$15 MeV,
provide a CP conserving phase which is effectively
maximal ($\pi/2$). Interference of the resonating amplitude with a direct
$B$ decay amplitude going through $b\to u\ubar d$, carrying a different CKM
phas
and leading to the same hadronic final states, creates a large CP asymmetry. We
will show  that due to this different and unusual mechanism of CP violation, CP
asymmetries in these charged $B$ decay processes are much larger than
in $B^{\pm}\to\psi\pi^{\pm}$, and are likely to reach a level of 10$\%$ or
more.
Furthermore, high statistics data may allow separate measurements of the
resonance amplitude and of the direct amplitude which acts as a background.
This
would allow determining the CKM phase  $\gamma={\rm arg}(V^*_{ub})$.

Resonance width effects in charged $B$ decay asymmetries were
studied recently
[6]. The leading effect was that of the interference of two (or more)
intermediate kaon resonances decaying to the same final states.
Interference between a resonant and a nonresonant
amplitude was already used as a CP violating mechanism in top quark decays,
where the W width is the source of a CP-even phase [7].

Let us describe in quite general terms the mechanism of CP violation in
$B^+\to X_c\pi^+$, in which the charmonium state $X_c=\eta_c$ or $\chi_{c0}$
decays to one of the above hadronic final states $h$, where for
instance $h=\rho\rho$ or $\pi^+\pi^-, K^+K^-$, respectively. For
simplicity, we will consider only two body and quasi two body $X_c$ decays.
In this case the $B^+$ decay distribution can be described in terms of $s$, the
center-of-mass energy-squared of the hadrons $h$, and $\theta$, the angle
between the $B$ momentum and the momentum of one of the two $X_c$ decay
products
in the $h$ center-of-mass frame. In a straightforward generalization to
multi-body decays, $\theta$ is replaced by several kinematical variables. We
denote by $a_1$ the weak decay amplitude of $B^+$ to $X_c\pi^+$ and by $a_2$
the
$X_c$ decay amplitude to $h$. The resonance amplitude, which also includes
a Breit-Wigner form for the intermediate $X_c$ state, is given by
$$
R(s)\equiv A(B^+\to X_c\pi^+\to
h\pi^+)=a_1a_2{\sqrt{\Gamma m}\over (s-m^2)+i\Gamma m} ~.\eqno(1)
$$
$m$ is the $X_c$ mass and $\Gamma$ is its width. To calculate the contribution
of the interference of this amplitude with another amplitude to a CP asymmetry,
subtraction of the partial decay width into $h$ is required by CPT [8]. Since
this partial width is very small relative to $\Gamma$, ${\cal O}(1\%)$ in our
cases of intererst, it will be neglected. Note that $R$ does not depend on
$\theta$ since $X_c$ is spinless. The CKM phase of $a_1$, given by  ${\rm
arg}(V^*_{cb}V_{cd})$, vanishes in the standard convention which we use. Hence
$a_1a_2$ is taken to be real. Final state phases due to rescattering from
other intermediate states will be absorbed into the strong phase of the direct
amplitude to be discussed below. (Only the relative phase is relevant).

For convenience, we will normalize the decay rate of $B^+\to X_c\pi^+\to
h\pi^+$ by the total $B$ decay rate
$$
{1\over\Gamma_B}{d^2\Gamma({\rm resonance})\over dsdz}=\vert R(s)\vert
^2~,~~~z\equiv\cos\theta~,\eqno(2)
$$
such that $a_1a_2$ is given by the product of the corresponding decay
branching ratios:
$$
2\pi\thinspace a^2_1 a^2_2={\rm BR}(B^+\to X_c\pi^+){\rm BR}(X_c\to h)~.
\eqno(3)
$$

As mentioned earlier, the $B^+$ decay amplitude into the final state $h\pi^+$,
at $s=m^2$, consists also of a direct decay term ($D$) induced by
$\bbar\to\ubar u d$ carrying a CKM phase $\gamma={\rm arg}(V^*_{ub})$.
We neglect a small contribution from penguin amplitudes [1]. The
possible slight  $s$-dependence of $D$ around $s=m^2$ will be neglected.
In general this amplitude depends on the angle $\theta$. The direct amplitude,
for $s\approx m^2$, is given by
$$
D(s\approx m^2, z)\equiv A(B^+\to
h\pi^+)={d(z)\over m_B}e^{i\gamma}e^{i\delta}~,\eqno(4)
$$
where $d(z)$ is real and  $\delta$ is a final
state interaction phase.
$d(z)$ can be decomposed into contributions from different spin-parity $h$
states:
$$
d(z)=\sum_{J^P}d_{(J^P)}(z)~,\eqno(5)
$$
in which an $S$-wave ($J=0$), for instance, corresponds to a constant term.
The direct decay rate will also be normalized by the total $B$ decay rate
such that $d(z)$ becomes dimensionless
$$
{1\over\Gamma_B}{d^2\Gamma({\rm direct}, s\approx m^2)\over
dsdz}=\vert D(s\approx m^2, z)\vert^2={d^2(z)\over m^2_B}~.\eqno(6)
$$

The $B^+\to h\pi^+$ amplitude at $s\approx m^2$, is given by a coherent
sum of the resonance amplitude $R$ and the direct amplitude $D$
$$
{1\over \Gamma_B}{d^2\Gamma^{(+)}\over dsdz}=\vert
R(s)+D(s\approx m^2, z)\vert^2~.\eqno(7)
$$
The corresponding amplitude for $B^-\to h\pi^-$ is obtained simply by changing
the sign of the weak phase $\gamma$ in $D(z)$. The difference and the sum of
$B^+$ and $B^-$ differential decay rates, {\it integrated symmetrically around
$s=m^2$}, say from $s=(m-2\Gamma)^2$ to $s=(m+2\Gamma)^2$, are given by
$$
{1\over\Gamma_B}({d\Gamma^{(+)}\over dz}-{d\Gamma^{(-)}\over
dz})\approx -12a_1a_2d(z){\sqrt{\Gamma m}\over m_B}\cos\delta\sin\gamma~,
$$
$$
{1\over\Gamma_B}({d\Gamma^{(+)}\over dz}+{d\Gamma^{(-)}\over
dz})\approx
6a^2_1a^2_2+16d^2(z){\Gamma m\over
m^2_B}-12a_1a_2d(z){\sqrt{\Gamma m}\over m_B}\sin\delta\cos\gamma~.\eqno(8)
$$

The partial rate asymmetry
$$
A\equiv {\Gamma^{(+)}-\Gamma^{(-)}\over\Gamma^{(+)}+\Gamma^{(-)}}~\eqno(9)
$$
requires an integration of (8) over $z$. In the numerator the resonance
amplitude interferes only
with a component of the direct amplitude corresponding to the hadronic system
$h$ with the charmonium $J^P$ quantum numbers.
Therefore, only one
$J^P$ term of $d(z)$ contributes , $0^-$ for
$\eta_c$, and $0^+$ for $\chi_{c0}$. In the
denominator we will assume for now that the resonance contribution to the
decay rate is much larger than the direct contribution over the resonance
region, $a^2_1a^2_2/\Gamma m\gg  d^2(z)/m^2_B$. We will comment on
corrections to this approximation when estimating the two contributions for the
relevant decays. Generally, the asymmetry depends on the phase
$\delta$ caused by rescattering effects in $D$ and in $R$, other than
due to the $X_c$ width. The dominant term in the
asymmetry is proportional to $\cos\delta$. Assuming that $\delta$
is small, which motivated our search for large resonance width effects in
the first place, we take $\cos\delta\approx 1$. We find
$$ A\approx -2({d_{(0^P)}\over
a_1a_2}){\sqrt{\Gamma m}\over m_B}\sin\gamma~.\eqno(10)
$$

Eq.(10) is our central general result. The asymmetry is given in
terms of twice the ratio of magnitudes of the direct and resonance amplitudes
at
$s=m^2$ times  $\sin\gamma$. Usually, an asymmetry contains also a sine of
a CP-conserving phase. {\it In our case, in which interference occurs between
th
resonance amplitude and the direct decay amplitude corresponding to the
background process at $s=m^2$, the strong phase difference is maximal, i.e.
$\pi/2$}.

The resonance amplitude $a_1a_2$ is given in (3) in terms of a product
of measurable branching ratios. The $0^P$ direct amplitude $d_{(0^P)}$ can be
obtained by a partial-wave analysis of the $z$-distribution slightly off
the resonance. Thus, we find
$$
\vert A\vert\approx 2\sqrt{\pi\Gamma m}\sqrt{{{1\over\Gamma_B}{d\Gamma({\rm
direct}, s\approx m^2, 0^P)\over ds}\over{\rm BR}(B^+\to X_c\pi^+){\rm
BR}(X_c\to h)}} \sin\gamma~,\eqno(11)
$$
where $d\Gamma({\rm direct}, s\approx m^2, 0^P)/ds$ is the $0^P$ contribution
to the differential decay rate slightly off the resonance.
This expression of the asymmetry can be used to determine the weak phase
$\gamma$ from measurable quantities.

To estimate the asymmetry, let us relate $d_{(0^P)}$, the $0^P$
direct amplitude, to a measurable integrated quantity. $d(z)$, the
total direct amplitude at $s\approx m^2$, may be estimated from
the $s$-integrated $B^+\to h\pi^+$ nonresonance branching ratio, ${\rm
BR}(B^+\to h\pi^+)_{\rm nonresonance}$. This branching ratio corresponds to $h$
systems which do not originate in other $s$-channel resonances.

In order to integrate over $s$ and $z$ the non-resonance differential decay
rate
$\vert  D(s, z)\vert^2$, which acts as a background, we will have to make an
assumption about its $s$-dependence. Using the variable $z$ (instead of the
usual second invariant square-momentum) introduces an extra $s$-dependent
factor
into  $D(s, z)$ relative to $D_{\rm inv.}(s, z)$, which is up to a constant
factor the commonly used invariant amplitude [2]:
$$
\vert D(s, z)\vert^2=\Phi(s)\vert D_{\rm inv.}(s, z)\vert^2~,
$$
$$
\Phi(s)\equiv \sqrt{1-{4m^2_0\over s}}(1-{s\over m^2_B})~.\eqno(12)
$$
We consider the case in which the
two hadrons in $h$ have equal masses $m_0$ and we set
$m_{\pi}\approx 0$. We will assume that the nonresonance invariant amplitude,
$D_{\rm inv.}$, is
approximately independent of $s$ over the entire range $4m^2_0\leq s\leq
m_B^2$:
$$
\vert D(s, m^2)\vert^2\approx {\Phi(s)\over\Phi(m^2)}{d^2(z)\over
m^2_B}~.\eqno(13)
$$
By integrating (6) over $s$ and $z$ we find
$$
{\rm BR}(B^+\to h\pi^+)_{\rm nonresonance}\approx
{I_{\Phi}\over 2\Phi(m^2)}\int_{-1}^1 d^2(z)dz={I_{\Phi}\over 2\Phi(m^2)}
\sum_{J^P}\int_{-1}^1 d^2_{(J^P)}(z)dz~,\eqno(14)
$$
where
$$
I_{\Phi}({m^2_0\over m^2_B})\equiv {2\over
m^2_B}\int_{4m^2_0}^{m^2_B}\Phi(s)ds~\eqno(15)
$$
is a standard 3-body phase space factor, which is very close
to one for $m_0=m_{\pi}$ and has a value of 0.68 for $m_0=m_{\rho}$.

To estimate the relative contribution of $d_{(0^P)}$ to the right-hand-side of
(14), one must apply model-dependent considerations.
We use
qualitative arguments, based on spin counting and on a partial wave analysis
for
the pion in $B^+\to h\pi^+$, which may be emitted from a very close distance to
the $b$ quark up to a typical hadronic distance away from it. This leads to a
suppression factor $f(0^P)$ of the $0^P$ decay rate relative to the total
direct
rate. This factor depends on the case under consideration and involves large
uncertainties. For instance, for $h=\pi^+\pi^-, K^+K^-$ in which $J=0$ requires
$P=+1$, we find $f(0^P)=0.07-0.7$, whereas for the cases $h=\rho\rho,
K\Kbar\pi$
where both parities are allowed in a $J=0$ state, the suppression may be
stronger. For definiteness, we use the above range for $f(0^P)$. Eq.(14) then
leads to
$$
d(0^P)\approx\sqrt{f(0^P){\Phi(m^2)\over I_{\Phi}}{\rm BR}(B^+\to h\pi^+)_{\rm
nonresonance}}~,
$$
$$
f(0^P)=0.07-0.7~. \eqno(16)
$$
Since the above limits on $f(0^P)$ correspond to extreme assumptions, it seems
to us that central values are more likely.
Using (3)(10)(16) we find
$$
\vert A\vert \approx \sqrt{f(0^P){\Phi(m^2)\over I_{\Phi}}}{\sqrt{8\pi\Gamma
m}\over m_B}\sqrt{{{\rm BR}(B^+\to h\pi^+)_{\rm nonresonance}\over
{\rm BR}(B^+\to X_c\pi^+){\rm BR}(X_c\to h)}}\sin\gamma~.\eqno(17)
$$

Let us estimate the asymmmetry under typical relevant circumstances. We will
use $\eta_c$ and $\chi_{c0}$ decay modes with branching ratios at a level of
1$\%$ [2]:
$$
{\rm BR}(X_c\to h) \sim 10^{-2}~.\eqno(18)
$$
The two $B^+$ decay branching ratios in (17) will evidently be known
before an asymmetry can be measured. We use the following value for the decay
branching ratio into $X_c\pi^+$:
$$
{\rm BR}(B^+\to\eta_c(\chi_{c0})\pi^+)\approx\vert{V_{cd}\over V_{cs}}\vert^2
{\rm BR}(B^+\to\eta_c(\chi_{c0})K^+)
$$
$$
\sim
\vert{V_{cd}\over V_{cs}}\vert^2
{\rm BR}(B^+\to\psi(\chi_{c1})K^+)\sim 5\times 10^{-5}~.\eqno(19)
$$
The branching ratios of $B^+\to\psi(\chi_{c1})K^+$ have been measured [2].
The common replacement $\pi^+\leftrightarrow K^+$ with corresponding CKM
factors is also justified by the recent measurement of $B^+\to\psi\pi^+$ [4].
Recent theoretical estimates of $\Gamma(B^+\to\eta_c K^+)/\Gamma(B^+\to
\psi K^+)$ [9] seem to indicate that
${\rm BR}(B^+\to\eta_c\pi^+)$ may be larger than (19) by about a factor 1.6 or
more.

The branching ratios of decays into nonresonant $h\pi^+$ states may vary
somewhat from case to case. We use as a characteristic value:
$$
{\rm BR}(B^+\to h\pi^+)_{\rm nonresonance}\sim 10^{-5}~.\eqno(20)
$$
This represents typical branching ratios of low multiplicity processes of
the type $b\to u\ubar d$, such as $B\to\pi\pi$ for which some evidence
already exists [10],
$B\to \pi\rho$, $B\to\pi\pi\pi$,  $B\to \pi\pi\rho$, $B\to
\pi K\Kbar$, etc. Branching ratios at this level were calculated for the above
two body ($\pi\pi$) and quasi two body ($\rho\pi$) decays by assuming
factorization [11].  Similar or even larger branching ratios are expected when
a
nonresonating pion is added to the final state, since at the high $B$ mass it
is
easy to fragment an additional pion. There is supporting evidence for this
behavior in $D$ decays, where ${\rm BR}(D^+\to \pi^+\pi^+\pi^-)_{\rm
nonresonant}\approx {\rm BR}(D^+\to \pi^+\pi^0)$ [2].  A statistical model for
the pion multiplicity as function of the available energy [12] predicts that in
$B$ decays the decay rate into three nonresonating pions should be larger than
for two pions.

Using the above values of branching ratios and central experimental values for
the $\eta_c$ and $\chi_{c0}$ masses and widths [2], we find from (17) similar
asymmetries for two representative processes,
$B^{\pm}\to (\rho^+\rho^-)_{\eta_c}\pi^{\pm}$
and $B^{\pm}\to (\pi^+\pi^-)_{\chi_{c0}}\pi^{\pm}$:
$$
\vert A\vert\sim 0.7\sqrt{f(0^P)}\sin\gamma~.\eqno(21)
$$
This is a rather large asymmetry for the
presently allowed values of $\gamma$, $0.3\leq\sin\gamma\leq 1$ [1]
and for the values of $f(0^P)$ in (16). Larger
asymmetries are obtained for larger values of ${\rm BR}(B^+\to h\pi^+)_{\rm
nonresonance}$ and for smaller values of ${\rm BR}(B^+\to X_c\pi^+)$ and ${\rm
BR}(X_c\to h)$.

We remind the reader that when deriving (10) we neglected the second term
in the right-hand-side of the lower eq.(8). The large asymmetry (21) indicates
that this direct contribution over the resonance region cannot be neglected. In
fact, for the interval of $s$ used to obtain eq.(8), it reduces the numerical
coefficient in (21) to about 0.5. This also affects eqs.(10) and (11) in a
similar manner. This correction can be made smaller by integrating over a
narrower range around the resonance.

The above estimated CP asymmetry applies to rather rare decay
processes, $B^+\to (h)_{s\approx m^2}\pi^+$, which have branching ratios ${\cal
B}$ of about
$$
{\cal B}\equiv {\rm BR}(B^+\to X_c\pi^+){\rm BR}(X_c\to h)\sim 5\times
10^{-7}~.\eqno(22)
$$
The number of charged $B$'s required for an observation of
such an asymmetry at a $3\sigma$ level is $N\approx 10({\cal B}A^2)^{-1}$,
which
depends only on  ${\rm BR}(B^+\to h\pi^+)_{\rm nonresonance}$ and not on ${\cal
B}$ itself:
$$
N\approx {10\over \sin^2\gamma}({m^2_B\over 8\pi\Gamma
m})({I_{\Phi}\over\Phi(m^2)}) ({0.7\over 0.5})^2({1\over f(0^P)})({1\over {\rm
BR} (B^+\to h\pi^+)_{\rm nonresonance}})~.\eqno(23)
$$
The last factor is likely to be smaller than $10^5$ for favorable decay modes,
such as $h= K\Kbar\pi, \pi^+\pi^-K^+K^-$ (for $\eta_c$) and $h=\pi^+\pi^-,
K^+K^-$ (for $\chi_{c0}$). The $0^P$ suppression
factor, for which a range of values was given in (16), is the most uncertain
one
and depends on the decay mode under consideration. Putting all numbers
together, we see that typically, for $\sin\gamma\sim 1$, about $10^8-10^9$
$B$'s
are needed to observe an asymmetry. For favorable cases, in which the $0^P$
suppression is weak (corresponding to a background that is flat in the
variable $z$) and in which the nonresonance branching ratio is large, fewer $B$
mesons may be required.

It is also possible to define another measurable CP violating quantity,
which does not involve the $0^P$ suppression factor, requiring however
measurement of the angular distributions
$d\Gamma^{(\pm)}/dz$. Denoting the asymmetry in $z$-distributions by $a(z)$
$$
a(z)\equiv {{d\Gamma^{(+)}\over dz}-{d\Gamma^{(-)}\over dz} \over
{d\Gamma^{(+)}\over dz}+{d\Gamma^{(-)}\over dz}}\approx
-2({d(z)\over a_1a_2})\sin\gamma~,\eqno(24)
$$
we define
$$
{\cal A}\equiv \sqrt{{1\over 2}\int^1_{-1} a^2(z)dz}~.\eqno(25)
$$
Measurement of ${\cal A}$ can be used to determine the weak phase $\gamma$ from
a relation similar to (11), however without requiring a partial-wave analysis:
$$
{\cal A}\approx 2\sqrt{\pi\Gamma m}\sqrt{{{1\over\Gamma_B}{d\Gamma({\rm
direct}, s\approx m^2)\over ds}\over{\rm BR}(B^+\to X_c\pi^+){\rm BR}(X_c\to
h)}} \sin\gamma~.\eqno(26)
$$
Using the approximation (13)(14) one obtains
$$
{\cal A}\approx \sqrt{{\Phi(m^2)\over I_{\Phi}}}{\sqrt{8\pi\Gamma m}\over
m_B}\sqrt{{{\rm BR}(B^+\to h\pi^+)_{\rm nonresonance}\over
{\rm BR}(B^+\to X_c\pi^+){\rm BR}(X_c\to h)}}\sin\gamma~.\eqno(27)
$$
Very large values, ${\cal A}\sim 0.5\sin\gamma$ (see (21) and discussion
below), are expected to be measured for this quantity.

We note that in the analogous Cabibbo-allowed decays $B^{\pm}\to (h)_{s\approx
m^2} K^{\pm}$, though the rates are larger than in
$B^{\pm}\to (h)_{s\approx m^2}\pi^{\pm}$ by a factor $\vert
V_{cs}/V_{cd}\vert^2$, the
asymmetries  are correspondingly smaller and harder to observe.  Finally, our
entire analysis applies generally to $B^+\to \eta_c(\chi_{c0})X^+$, where $X^+$
is any hadronic state made from $u\dbar$, such as $\rho^+, \pi^+\pi^0$, etc.
One
may also consider the semi-inclusive processes $b\to d\eta_c(\chi_{c0}),~
\eta_c(\chi_{c0})\to h$, similar to $b\to d \psi$ [5]. Their folded branching
ratios are expected to be about $5\times 10^{-6}$, an order of magnitude larger
than the exclusive branching ratios. Their asymmetries are as large as in the
exclusive decays. The observation of such asymmetries would be easier if
measurement of $b\to d\eta_c(\chi_{c0})$ were possible, in spite of the about
twenty times larger background from $b\to s\eta_c(\chi_{c0})$.

\vskip 0.5truecm
We thank J. Goldberg, B. Kayser, H. J. Lipkin and S. Stone for useful
discussions. This work was supported in part by the German-Israeli Foundation
for Scientific Research and Development, by the Fund for Promotion of
Research at the Technion and by the Natural Sciences and Engineering Research
Council of Canada.
\vfill\eject

\centerline{\bf References}
\vskip0.5truecm
\noindent
[1] Y. Nir and H. R. Quinn in {\it B Decays}, ed. S. Stone (World Scientific,
1994), p. 520; I. Dunietz, {\it ibid.}, p. 550; M. Gronau, {\it Proceedings of
Neutrino 94, XVI International Conference on Neutrino Physics and
Astrophysics},
Eilat, Israel, May 29 - June 3, 1994, eds. A. Dar, G. Eilam and M. Gronau,
{\it Nucl. Phys. (Proc. Suppl.)} {\bf B38}, 136 (1995).
\hfil\break
[2] Particle Data Group, {\it Phys. Rev.} {\bf
D50}, 1173 (1994).
\hfil\break
[3] H. Yamamoto, HUTP-94/A006 (1994).
\hfil\break
[4] CLEO Collaboration, J. P. Alexander {\it et al.}, {\it Phys. Lett.} {\bf
B341}, 435 (1995).
\hfil\break
[5] I. Dunietz, {\it Phys. Lett.} {\bf B316}, 561 (1993); I. Dunietz and J. M.
Soares, {\it Phys. Rev.} {\bf D49}, 5904 (1994); J. M. Soares, HEP-PH-9404336.
\hfil\break
[6] D. Atwood and A. Soni, {\it Z. Phys.}, {\bf C64} 241 (1994) and
{\it Phys. Rev. Lett.} {\bf 74}, 220 (1995); D. Atwood, G. Eilam, M. Gronau and
A. Soni, {\it Phys. Lett.} {\bf B341}, 372 (1995).
\hfil\break
[7] G. Eilam, G. Hewett and A. Soni, {\it Phys. Rev. Lett.} {\bf 67}, 1979
(1991); {\it ibid.} {\bf 68}, 2103 (1992); D. Atwood, G. Eilam, A. Soni, R. R.
Mendel and R. Migneron, {\it Phys. Rev. Lett.} {\bf 70}, 1364 (1993) and
references therein.
\hfil\break
[8] J-M. Gerard and W-S. Hou, {\it Phys. Rev. Lett.} {\bf 62}, 855 (1989);
L. Wolfenstein, {\it Phys. Rev.} {\bf D43}, 151 (1991).
\hfil\break
[9]  N. G. Deshpande and J.
Trampetic, {\it Phys. Lett.} {\bf B339}, 270 (1994). M. R. Ahmady and R. R.
Mendel,  HEP-PH-9401327, to
be published in {\it Z. Phys.} {\bf C};~ M. Gourdin,  Y. Y. Keum and X. Y.
Pham, HEP-PH-9409221.
\hfil\break
[10] CLEO Collaboration, M. Battle {\it et al.}, {\it Phys. Rev. Lett.} {\bf
71}, 3922 (1993).
\hfil\break
[11] M. Bauer, B. Stech and W. Wirbel, {\it Z. Phys.} {\bf C34}, 103
(1987); M. Wirbel, {\it Prog. Part. Nucl. Phys.} {\bf 21}, 33 (1988).
\hfil\break
[12] C. Quigg and J. L. Rosner, {\it Phys. Rev.} {\bf D16}, 1497 (1977); {\it
ibid.} {\bf D17}, 239 (1978); A. Ali, J. G. Korner, G. Kramer and J. Willrodt,
{\it Z. Phys.} {\bf C2}, 33 (1979).
\hfil\break
\bye